\documentclass[
 aps, prd,
 amsmath,amssymb,
 reprint
]{revtex4-2}

\usepackage{graphicx}% Include figure files
\usepackage{dcolumn}% Align table columns on decimal point
\usepackage{bm}% bold math
\usepackage{multirow}
\usepackage[utf8]{inputenc}
\usepackage[T1]{fontenc}
\usepackage{mathptmx}
\usepackage{etoolbox}

\graphicspath{ {./fig/} }

\begin{document}

\title{Degaussing Procedure and Performance Enhancement by Low-Frequency Shaking of a 3-Layer Magnetically Shielded Room}
\author{Fabian Allmendinger}
 \email{allmendinger@physi.uni-heidelberg.de}
 \affiliation{Physikalisches Institut, Universit\"at Heidelberg} 
\author{Benjamin Brauneis}
 \affiliation{Physikalisches Institut, Universit\"at Heidelberg}
\author{Werner Heil}%
 \affiliation{Institut f\"ur Physik, Universit\"at Mainz}
\author{Ulrich Schmidt}
 \affiliation{Physikalisches Institut, Universit\"at Heidelberg}

\date{\today}

\begin{abstract}
We report on the performance of a Magnetically Shielded Room (MSR) intended for next level $^3$He/$^{129}$Xe co-magnetometer experiments which require improved magnetic conditions. The MSR consists of three layers of Mu-metal with a thickness of 3~mm each, and one additional highly conductive copper-coated aluminum layer with a thickness of 10~mm. It has a cubical shape with an walk-in interior volume with an edge length of 2560~mm. An optimized degaussing (magnetic equilibration) procedure using a frequency sweep with constant amplitude followed by an exponential decay of the amplitude will be presented. The procedure for the whole MSR takes 21 minutes and measurements of the residual magnetic field at the center of the MSR show that $|B|<$~1nT can be reached reliably. The chosen degaussing procedure will be motivated by online hysteresis measurements of the assembled MSR and by Eddy current simulations showing that saturation at the center of the Mu-metal layer is reached. Shielding Factors can be improved by a factor $\approx 4$ in all directions by low frequency (0.2~Hz), low current (1~A) shaking of the outermost Mu-metal layer.
\end{abstract}

\maketitle

%%%%%%%%%%%%%%%%%%%%%%%%%%%%%%%%%%%%%%%%%%%%%%%%%%%%%%%%%%
%%%%%%%%%%%%%%%%%%%%%%%%%%%%%%%%%%%%%%%%%%%%%%%%%%%%%%%%%%%
\section{\label{sec:Introduction}Introduction}
Magnetically Shielded Rooms (MSRs) have become increasingly important both in fundamental physics and applied research due to technological developments in the field of large-volume, and thus walk-in, accessible shielding rooms. Residual magnetic fields of less than 1~nT with field gradients below 10~pT/cm with strong damping of electromagnetic distortions over a wide range of frequencies can be achieved in a volume of $>1~\text{m}^3$.
There are various applications in applied research like Biomagnetism~\cite{Fedele} or ultra-low field nuclear magnetic resonance~\cite{Volegov}. Magnetic shielding is also crucial for experiments in fundamental physics, prominently in searches for Electric Dipole Moments (EDMs) of elementary particles or composite particles like the neutron~\cite{Lauss} or the neutral $^{199}$Hg atom~~\cite{Graner}.\\
The MSR described in this work is intended for the $^3$He/$^{129}$Xe co-magnetometer experiment, a high precision experiment at low energies which can address a variety of fundamental questions associated with symmetry violations in nature~\cite{Heil}. Worth mentioning here are: the measurement of the \textit{CP}-violating EDM of the $^{129}$Xe atom~\cite{Allmendinger}, looking for a violation of Lorentz Invariance~\cite{Allmendinger2}, and searching a spin-dependent \textit{P}- and \textit{CP}-violating nucleon-nucleon interaction mediated by Axions or axion-like particles~\cite{Tullney}. In short, the measurement principle is: Two co-located spin samples (hyperpolarized $^3$He and $^{129}$Xe gas) are used as a sensitive probe for these non-magnetic spin interactions~\cite{Gemmel}. Their Larmor frequencies measured by low-temperature SQUIDs are compared while the effect under investigation is varied by, for example, periodically inverting an applied electric field leading to a corresponding modulation of the Larmor frequency of $^{129}$Xe for a non-zero EDM.\\
Next level $^3$He/$^{129}$Xe co-magnetometer experiments require improved magnetic conditions as follows: Firstly, statistical uncertainties are anti-proportional to the Signal-to-Noise Ratio (SNR) which can be increased by suppressing the effect of external noise sources. A common parameter to describe the performance of magnetic shields is the Shielding Factor (SF), the ratio of the magnetic flux density $B$ measured at the center of the shielded volume and the magnetic flux density without shielding at the same position. At the relevant frequency range ($\approx$ 1 to 20~Hz) shielding factors have to exceed 3000. As the shielding material in itself is a noise source due to Johnson noise~\cite{Romalis}, a larger distance (>1~m) of the the magnetometers to the shielding material is necessary. Secondly, measurement sensitivity is influenced by the stability and homogeneity of the magnetic field inside the MSR. Spin coherence times $T_2^*$ of several hours can be achieved only with low magnetic field gradients on the 10~pT/cm order of magnitude~\cite{Gemmel}. Statistical uncertainties are proportional to $(T_2^*)^{-3/2}$, so that measurement sensitivity benefits strongly from low gradients in a central volume (20x20x20~cm$^3$) containing the spin samples. Experience from previous experiments teaches that our measurement method using co-magnetometry works well, if the relative drift of the resulting inner magnetic field is less than $10^{-3}$ per hour. Since the guiding field is $\approx1$~µT, this corresponds to a drift of the residual field of less than 1~nT/h. As static magnetic field gradients can be further compensated by a gradient coil system inside the MSR, residual fields of less than 1~nT should be strived for to keep the resulting gradients and their potential temporal drifts sufficiently low.\\
It should be noted that the magnetic requirements for the $^3$He/$^{129}$Xe co-magnetometer experiment are more relaxed compared to neutron EDM experiments, e.~g., at PSI ~\cite{Abel, Lauss}, which need the specified residual field values and field homogeneity over a larger volume of 1x1x1~m$^3$ due to the larger EDM spectrometer size.\\
To meet these requirements, a magnetically shielded room was installed at Physikalisches Institut at Heidelberg, Germany. The design of this MSR, its degaussing procedure and resulting magnetic properties are the focus of this paper.\\
The outline of the work is as follows:\\
In the first part, specifications and design details of the magnetic shielding are given, followed by a description of the degaussing (demagnetization) procedure with its electronics setup and degaussing sequence. The next part covers the magnetic performance including residual field and shielding factors after proper degaussing without and with shielding factors enhancement by low-frequency shaking. The third part motivates the unusual choice of the degaussing parameters (especially the low frequency of 0.5~Hz) by means of an online hysteresis measurement of the assembled MSR and Eddy-current simulations. Finally, an outlook will be given pointing to future magnetic gradients measurements and an additional active residual field and gradients compensation. 
%%%%%%%%%%%%%%%%%%%%%%%%%%%%%%%%%%%%%%%%%%%%%%%%
%%%%%%%%%%%%%%%%%%%%%%%%%%%%%%%%%%%%%%%%%%%%%%%%%%%
\section{\label{sec:MSRdesign}Properties of the Magnetically Shielded Room}
\begin{figure}
    \centering
    \includegraphics[width=0.5\textwidth]{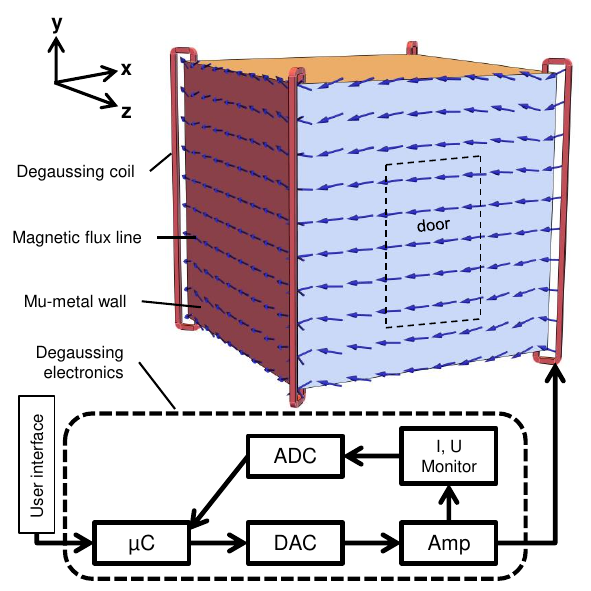}
    \caption{Degaussing system (schematic): Four coils (red) around the edges of a single Mu-metal layer are connected in series to a degaussing coil unit generating a closed magnetic flux loop (blue vectors) through four of the Mu-metal walls (result of a Radia simulation). In this example, the flux through the top and bottom walls is negligible.  Such a unit exists for every spatial direction for all three layers (9 in total, not shown). The degaussing electronics consists of a microcontroller and 20-bit digital-to-analog converter that generate the chosen waveform. After low-pass filtering, the signal is fed to a 4-quadrant power amplifier. Feeding back the digitized current and voltage output to the microcontroller allows for an online offset compensation. Not shown are the relays for connecting/switching the output of the Power Amplifier to the different degaussing coil units.
    }    \label{fig:degcoils}
\end{figure}

To attain the required magnetic conditions the usual approach of passive magnetic shielding was adopted: The sensitive experiment is enclosed by a high-permeability material which acts as a flux shunt for the external field that may vary in time (earth magnetic field, moving ferromagnetic objects like trucks, electric equipment). In this case, the MSR, manufactured by Vacuumschmelze~\cite{Vac1}, has a cubical shape and consists of three concentric layers of Mu-metal, a NiFe-alloy, with a thickness of 3~mm each. The edge lengths are 2965~mm, 2605~mm and 2560~mm, respectively. The construction of each of the three Mu-metal layers was as follows: Layers of thin (600~µm) Mu-metal sheets were laminated crosswise to form larger (up to 1300x1300~mm$^2$) and thicker (3~mm) modules (flat sheets and edge pieces). Low magnetic resistance between modules was achieved by overlapping areas. The thickness of the modules was reduced at overlapping areas to ensure a constant thickness of 3~mm throughout the Mu-Metal wall. This way, air gaps were avoided.\\
An additional highly conductive copper-coated aluminum layer with a thickness of 10~mm between the outer and middle Mu-metal layer serves as an electromagnetic shield (Eddy-current layer) for higher frequencies (including RF shielding).\\
All shielding layers are fixed to an aluminum beam support structure that also acts as a rigid mounting frame for all components of the experiment inside the MSR, thereby preventing or reducing vibrations of individual components relative to each other and the Mu-metal walls.\\
The whole MSR rests on a rigid platform build from aluminum beams and plates with a total height of 210~mm. The reasons are, firstly, to increase the distance from the bottom Mu-metal floor to the Lab floor which contains magnetized steel enforcement, and secondly, allowing future installments of active compensation coils of the ambient magnetic field.\\
The MSR is accessible through a door (clearance 2000~mm~x~1000~mm) that is placed at the center of the front wall, slightly shifted to the bottom to ensure level entrance to the inside. The door is opened and closed manually. Latching clamps are activated pneumatically and ensure a constant and well-defined pressure between the overlapping shielding material of wall and door, and thereby, good magnetic contact. Similarly, good electrical contact between the individual parts of the Eddy-current layer prevents a degrading shielding factor at higher frequencies.\\
Inside the MSR, a walkable wooden floor allows the manual installation of components. The wooden floor rests on the support structure. All components of the measurement setup inside the MSR (like coils, magnetometers) can be attached to the aluminum mounting frame. Therefore, the two inner Mu-metal layers have cutouts for 9 mounting points per wall including ceiling and floor, except the front wall, which has 6.\\
There are several RF-shielded ports and openings with a diameter between 60 and 160~mm that are used for ventilation, transfer of polarized gases etc. The ports are positioned closely to corners and edges of the MSR to minimize the disturbance of the magnetic field inside.\\
The MSR is equipped with degaussing coils consisting of 5 turns of 2.5~mm$^2$ copper wire around each of the 12 edges of each Mu-metal layer. There are cutouts at the corners of the respective Mu-metal layers for feeding through the coils and connecting wires. Four coils each are connected in series to a \textit{degaussing coil unit} generating a closed magnetic flux loop through four of the Mu-metal walls. Fig.~\ref{fig:degcoils} shows the magnetization vectors inside a Mu-metal layer generated by such a single degaussing coil unit (result of a simulation using Radia~\cite{Radia1,Radia2}).  The magnetization is homogeneous inside four walls, while the magnetic flux leakage into the remaining two walls (in our example the top and bottom walls) of the Mu-metal layer is negligible small. Usually all three spatial directions are degaussed successively, covering each wall twice. However, the flux loops using only two degaussing coil units cover all six walls already, which can be utilized in time-saving degaussing procedures (see Tab.~\ref{tab:sequences}). The other two Mu-metal layers have an identical coil configuration, so that there are in total 9 degaussing coil units. Each degaussing coil unit is connected to the external current source via twisted pair feed lines with an additional electric shielding. Whenever a degaussing coil unit is currently not in use, double switching relays break both connections to the amplifier to avoid feeding RF noise into the MSR.\\
The experiments coordinate system is indicated in Fig.~\ref{fig:degcoils}. The main axes are parallel to the Mu-metal walls with the x-axis left to right, y-axis bottom to top, and z-axis back to front wall (seen from outside the MSR, facing the door; the front wall contains the door). The origin of the coordinate system coincides with the center of the MSR.\\
There are three quadratic Helmholtz-like coil pairs for shielding factor measurement with 7 turns per single coil of 0.75~mm² copper wire along the outer edges of the MSR. The dimensions are approximately 3~m by 3~m with a spacing of 3~m. The hypothetical magnetic field produced by these coils at the center of the MSR if no Mu-metal was present can be calculated according to 
\begin{eqnarray} \label{eqn:coilfactors}
\begin{pmatrix}
B_\text{x}\\
B_\text{y}\\
B_\text{z}
\end{pmatrix}
=
\begin{pmatrix}
2.196&0&0\\
0&2.162&0\\
0&0&2.196
\end{pmatrix}
\cdot
\begin{pmatrix}
I_\text{x}\\
I_\text{y}\\
I_\text{z}
\end{pmatrix}
\cdot ~\text{µT/A}
\end{eqnarray}
where $I_\text{x}$, $I_\text{y}$ and $I_\text{z}$ are the coil currents.
\\

%%%%%%%%%%%%%%%%%%%%%%%%%%%%%%%%%%%%%%%%%%%%%%%%
%%%%%%%%%%%%%%%%%%%%%%%%%%%%%%%%
\section{\label{sec:Degaussing}Degaussing procedure}
The inner two Mu-metal walls of the MSR will be magnetized if exposed to stronger magnetic fields which is usually the case when the door of the MSR has been opened, for example. To reach reproducible magnetic conditions (residual field and shielding factors), degaussing, i.~e., the elimination of a remnant magnetization of the whole MSR is necessary. This is usually achieved by applying a decreasing sinusoidal current to the degaussing coil units. The following conditions must be met to effectively demagnetize: Firstly, the maximum peak applied magnetic field (or current) must be sufficiently high to saturate the magnetic material in every region of the shielding. Secondly, the amplitude decrease must be slow enough, so that consecutive maxima have a small difference (rule of thumb: 1\%), leading to a random orientation of magnetic domains with zero magnetization~\cite{Thiel1}. 
%%%%%%%%%%%%%%%%%%%%%%%%%%%
\subsection{\label{sec:DegaussingSetup}Degaussing setup}
The degaussing hardware consists of the following main building blocks as depicted in the bottom part of Fig.~\ref{fig:degcoils}: waveform generation, power amplifier, and degaussing coil units. The waveform is generated in software using a microcontroller which updates a digital-to-analog converter (DAC) with a resolution of 20 bits (Analog Devices AD5790) periodically with an update rate of 1~kHz. Specific degaussing waveforms are discussed in the following paragraph. After low-pass filtering (cutoff frequency $\approx200$~Hz) the signal is fed into a 4-quadrant power amplifier (T\"ollner TOE 7621-20,~\cite{Toellner}) with maximum voltage and current output of $\pm20$~V and $\pm16$~A. The output of the power amplifier is connected to the different degaussing coil units by double switching relays controlled by the microcontroller.\\
Note that there is no transformer in our setup which is often used to eliminate any current offset~\cite{Thiel1}. Here, we use a different approach: Feeding back the digitized current and voltage output of the power amplifier to the microcontroller allows for an online offset compensation in software, i.~e., slowly adjusting an offset to the DAC output to compensate offset errors further down the signal chain. The main reason lies in the unusually low degaussing frequency below 1~Hz (needed to effectively degauss the 3~mm thick Mu-metal walls, see Sec.~\ref{sec:Motivation}) that makes transformer coupling very unpractical. 
%%%%%%%%%%%%%%%%%%%%%%%%%%%%%%%
\subsection{\label{sec:DegaussingSequence}Degaussing waveforms and sequence}
\begin{figure}
    \centering
    \includegraphics[width=0.5\textwidth]{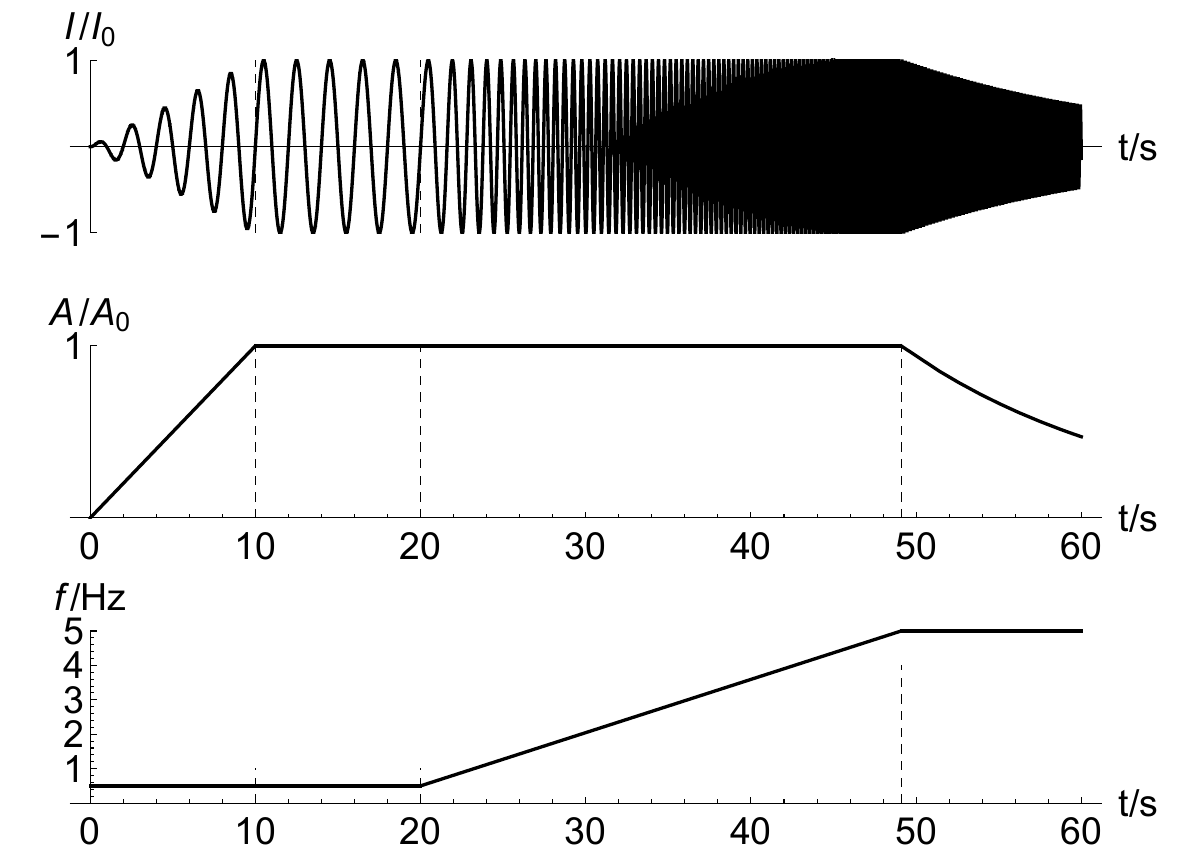}
    \caption{Typical degaussing waveform. Ratio of the current through the degaussing coil unit $I$ to the maximum current $I_0=16$~A as a function of time (top). The four sections (distinguished by vertical dashed lines) are best described as a combination of amplitude and frequency modulation with the normalized amplitude $A/A_0$ (middle) and frequency $f$ (bottom).
    }    \label{fig:waveform}
\end{figure}

The concept of using a microcontroller and a DAC allows for freely programmable degaussing waveforms including a combination of amplitude and frequency modulation. Several degaussing sequences have been tested and compared to each other with respect to shielding performance and total degaussing duration (see Tab.~\ref{tab:sequences}). In our case it proved useful to describe waveforms as a combination of amplitude and frequency modulation as depicted in Fig.~\ref{fig:waveform}. Waveforms are subdivided into four sections: In the first section (duration: a few seconds) the waveform is a sinusoidal signal with frequency $f_0$ and linearly increasing amplitude, followed by a second section (duration: a few seconds) with constant frequency and constant amplitude. In the third section ("sweep"), the frequency increases linearly from $f_0$ to $f_1$ within a period of $t_\text{sweep}$ while the amplitude stays constant. In the last section, the waveform is a sinusoidal signal with frequency $f_1$ and an exponentially decreasing amplitude, where $r$ gives the ratio of consecutive maxima. The fourth section lasts until the remaining amplitude is too small to be resolved by the 20-bit DAC.\\
As all three layers have to be demagnetized, the question arises which layer sequence and there in turn which ordering of the two or three flux loop directions should be used. We defined and tested  degaussing sequences as shown in Tab.~\ref{tab:sequences}. Here, we define the degaussing directions as follows: $y$-direction means the closed flux loop in the $x$-$z$-plane, i.~e., a magnetic flux through the left, back, right, front walls as depicted in Fig.~\ref{fig:degcoils}. Correspondingly, we have the $x$-direction with a closed flux loop in the $y$-$z$-plane, and the $z$-direction with a closed flux loop in the $x$-$y$-plane.\\
The impact of different degaussing waveforms and sequences on the MSR shielding performance (especially the residual field) will be covered in the next section.
\begin{table}
\centering
\begin{tabular}{|c||l|r|r|r|r||r|r|} 
 \hline
\multirow{2}{*}{\#}&\multirow{2}{*}{Sequence}&\multirow{2}{*}{$f_0$/Hz}&\multirow{2}{*}{$f_1$/Hz}&\multirow{2}{*}{$t_\text{Sweep}$/s}&\multirow{2}{*}{$r$}&\multirow{2}{*}{$T$~/min}&\multirow{2}{*}{$|B|$~/nT}\\
&&&&&&&\\
 \hline
 \hline
 \multirow{2}{*}{1}&1x-1y-1z-2x-2y-2z&0.5&0.5&0&0.992&\multirow{2}{*}{516}&\multirow{2}{*}{<1}\\
\cline{2-6}
&3x-3y-3z&0.5&0.5&0&0.992&&\\
\hline
\hline
\multirow{2}{*}{2}&1x-1y-2x-2y&0.5&10&100&0.986&\multirow{2}{*}{184}&\multirow{2}{*}{<1}\\
\cline{2-6}
&3x-3y-3z&0.5&0.5&0&0.986&&\\
\hline
\hline
\multirow{2}{*}{3}&1x-1y-2x-2y&0.5&10&100&0.986&\multirow{2}{*}{21}&\multirow{2}{*}{<1}\\
\cline{2-6}
&3x-3y-3z&0.5&10&100&0.986&&\\
\hline
\hline
\multirow{2}{*}{4}&1x-1y-2x-2y&0.5&10&40&0.986&\multirow{2}{*}{14}&\multirow{2}{*}{<2}\\
\cline{2-6}
&3x-3y-3z&0.5&10&40&0.986&&\\
\hline

\end{tabular}
\caption{Degaussing sequences, described by characteristic parameters:
sequence of layers and directions. Here, 1 is the outer, 2 the middle, and 3 the inner Mu-metal layer. See Sec.~\ref{sec:DegaussingSequence} for directions definitions. The degaussing waveforms are parameterized by: the initial frequency $f_0$, final frequency $f_1$, duration of the sweep section $t_\text{sweep}$, ratio $r$ of consecutive maxima during the decay section.  The total duration $T$ for the complete degaussing sequence of all layers is given, and the abs. value of the residual field in the center of the MSR as a measure of the degaussing quality. For comparison: The residual field in the closed MSR after the door was open for more than 10 minutes is $|B|\approx 10$~nT.  See Sec.~\ref{sec:PerformanceResidual} for a discussion of the results.}
\label{tab:sequences}
\end{table}
%%%%%%%%%%%%%%%%%%%%%%%%%%%%%%%%%%%%%%%%%%%%%%%%%%
%%%%%%%%%%%%%%%%%%%%%%%%%%%%%%%%%%%%%%%%%%%%%%%%%%
\section{\label{sec:Performance}Shielding performance measurement setup and results}
The magnetic shielding performance is usually characterized by 1) Shielding Factors (SFs), i.~e., the attenuation factors for external magnetic perturbations which generally are frequency and amplitude dependent, 2) the residual magnetic field inside the MSR (typically measured at the center) and its drift over time, and 3) magnetic field gradients in a volume of interest ~\cite{Sun1,Altarev1,Voigt1}. The measurements of these characteristic values are covered in the following sections.
\subsection{\label{sec:Magnetometer}Magnetometer gain and offset calibration}
\begin{figure}
    \centering
    \includegraphics[width=0.5\textwidth]{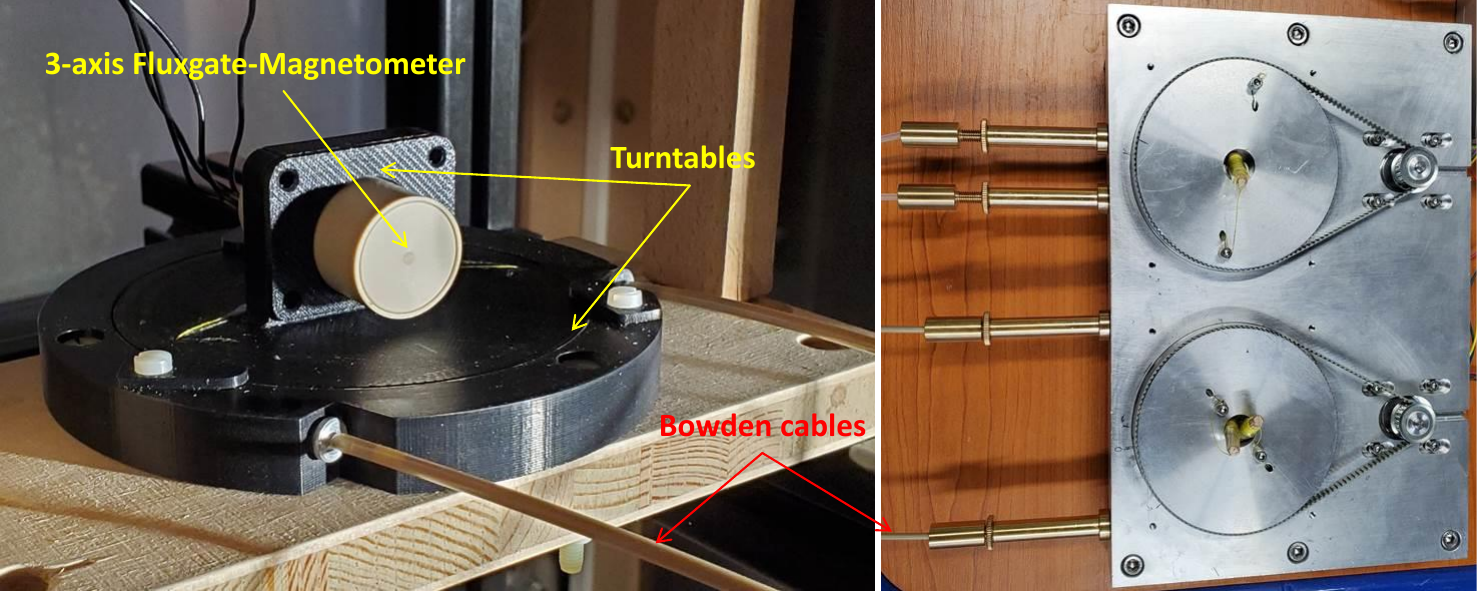}
    \caption{Rotating mechanism for the tri-axial Fluxgate magnetometer. Left: Non-magnetic 2-axis turntable inside the MSR, moved by Bowden cables. Right: Stepper motors outside the MSR actuate the Bowden cables.
    }    \label{fig:rot}
\end{figure}

Here, we use the following Fluxgate magnetometer models from Stefan Mayer Instruments~\cite{StefanMayer1}: the tri-axial model FLC3-70 with an amplitude noise density of $\rho=120~\text{pT}/\sqrt{\text{Hz}}$ at 1~Hz, and the single axis model FL1-100 with lower noise of $\rho=20~\text{pT}/\sqrt{\text{Hz}}$ at 1~Hz. We calibrated the magnetometers using a long solenoid with precisely known coil factor and a sinusoidal current (AC measurement, $f=1$~Hz). Gain drifts are small and of minor concern in the context of this paper. However, magnetometer offset (and drift thereof) have to be precisely known to measure small residual fields of order 1~nT reliably as this close to the sensitivity limit of the Fluxgates. The solution is: We monitor the offset by rotating the sensor on a non-magnetic 2-axis turntable, so that each axis of the sensor is inverted at least once. The sensor reading due to an actual magnetic field changes sign; however, the offset stays constant. Fig.~\ref{fig:rot} shows the mechanical setup with the non-magnetic 2-axis turntable made of plastic which can be turned by two sets of non-magnetic Bowden cables (poly-amid string in a PEEK resp. poly-amid cable housing). The part outside the MSR pulling the Bowden cables is actuated by two stepper motors. To reduce mechanical vibrations, the setup is rotated slowly; a 360$^\circ$ rotation takes typically 20~s. The sensor is first rotated by 360$^\circ$ in the horizontal plane, then by 360$^\circ$ around the symmetry axis of the cylindrical senor housing. A sinusoidal fit to the data gives the sensor offsets (for each sensor axis individually) and the actual magnetic field amplitudes. The order of magnitude of sensor offsets is 10~nT, with drifts of less than 0.5~nT/h. 

\subsection{\label{sec:PerformanceSF}Shielding Factors}
\begin{figure}
    \centering
    \includegraphics[width=0.5\textwidth]{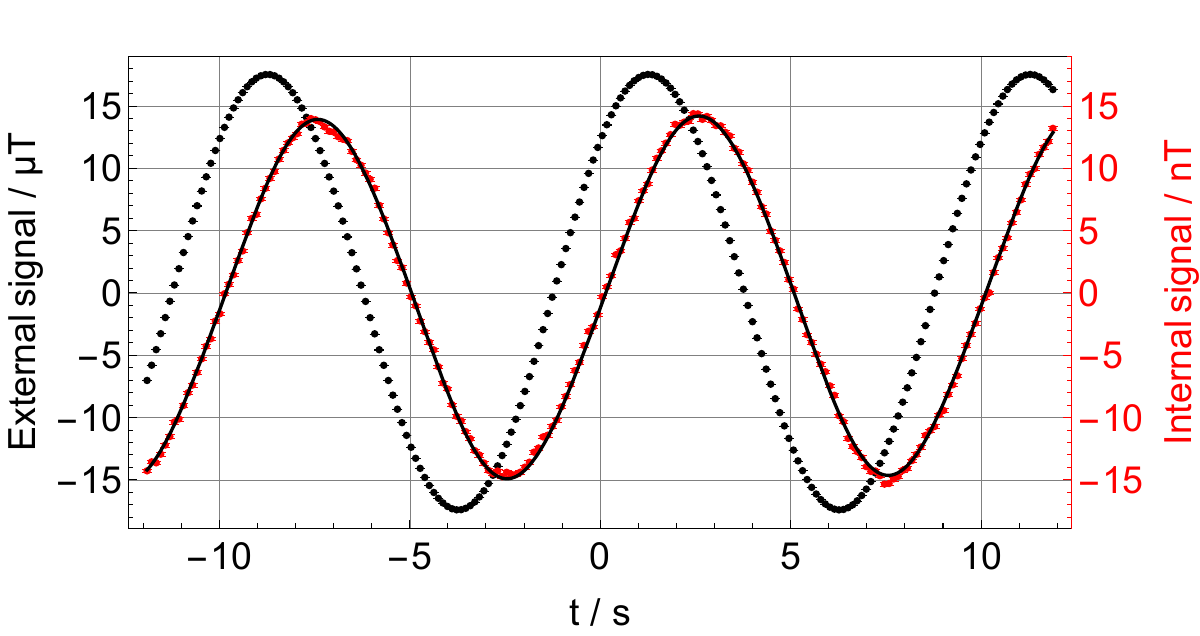}
    \caption{Shielding Factor measurement: External (black) and internal (red) signal amplitudes and fitted model to extract shielding factors and phases, here, for a large external amplitude of $17.5$~µT and $f=0.1$~Hz, z-direction. The SF is $\approx1250$ and the phase shift $\Delta\Phi=-1$~rad.
    }    \label{fig:SF1}
\end{figure}

\begin{figure}
    \centering
    \includegraphics[width=0.5\textwidth]{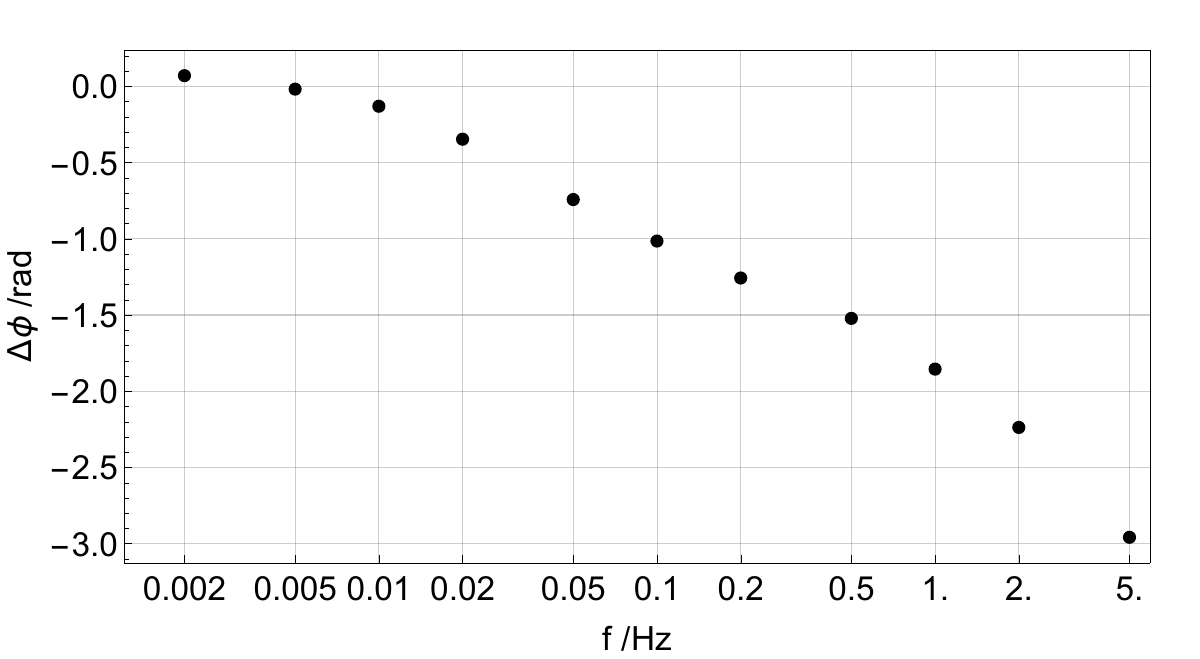}
    \caption{Phase shift of the internal signal with respect to the external signal as a function of frequency. The external amplitude was 1.75~µT, z-direction. No shaking was applied. Measurement uncertainties are smaller than the symbol size.
    }    \label{fig:phases}
\end{figure}

\begin{figure}
    \centering
    \includegraphics[width=0.5\textwidth]{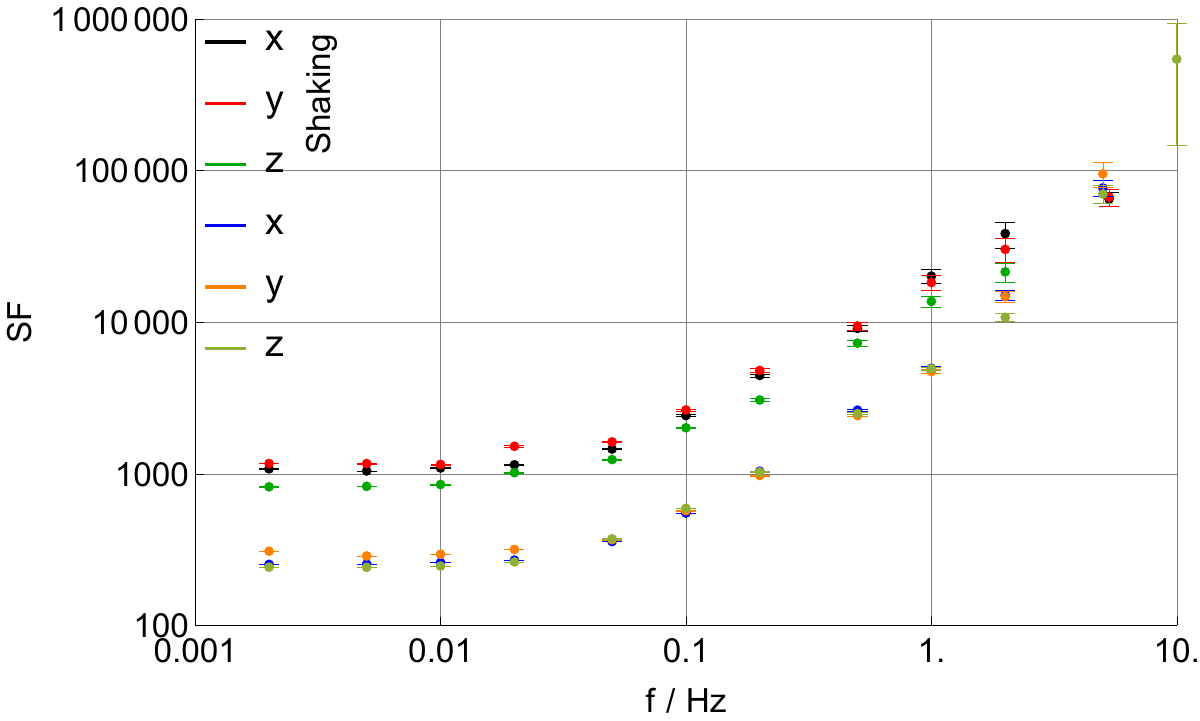}
    \caption{Shielding Factors as a function of frequency for different directions. The external amplitude was 1.75~µT. Data were taken without (bottom data points) and with shaking (top data points, shaking with frequency 5~Hz and current 4~A, $z$-direction, see Sec.~\ref{sec:shaking}).
    }    \label{fig:SF2}
\end{figure}

\begin{figure}
    \centering
    \includegraphics[width=0.5\textwidth]{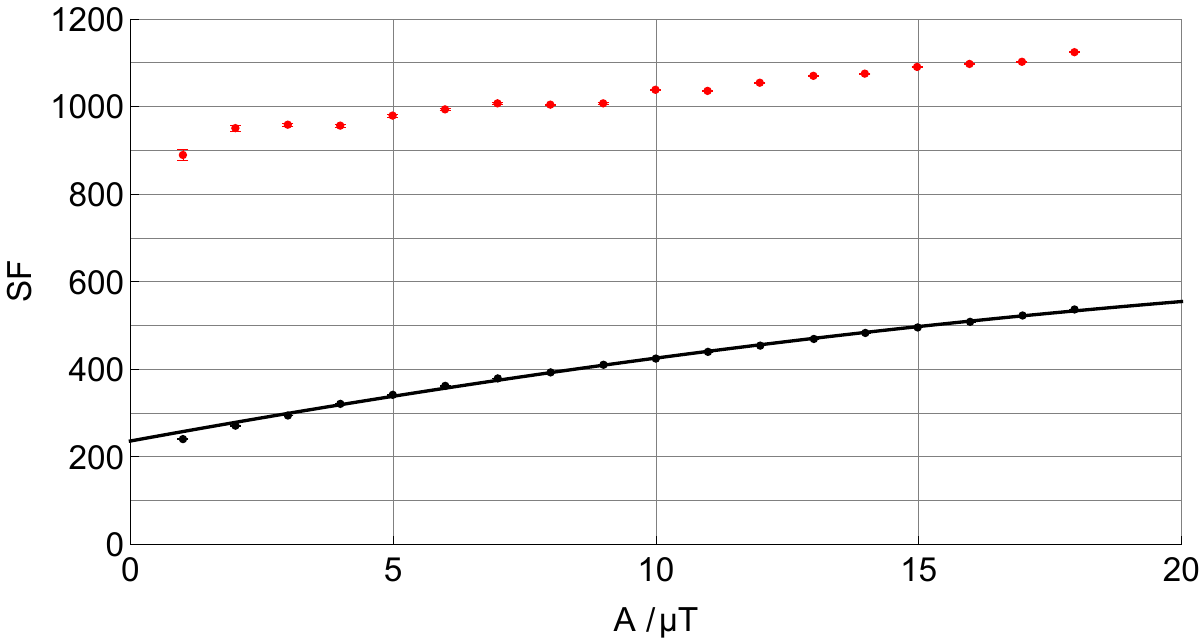}
    \caption{Shielding Factors as a function of the external amplitude without (black) and with (red) shaking (5~Hz, 4~A, $z$-direction). The signal frequency is $f=20$~mHz; in $z$-direction. Fit of a quadratic function $a+b\cdot A + c\cdot A^2$ to the measurement data without shaking yields $a=235.8 \pm 0.5$, $b=(21.94 \pm 0.06)$/µT and $c=(-0.301 \pm 0.004)$/µT$^2$.
    }    \label{fig:SF3}
\end{figure}

Shielding Factors have been determined separately for each main axis for a set of different frequencies and amplitudes using the following method: A signal generator and a 4-quadrant amplifier (T\"ollner model TOE 7621-40, max. voltage $\pm 40$~V, max. current $\pm 8$~A) control a current through one of the three Helmholtz-like coil pairs at the outer edges of the MSR corresponding to the $x$-, $y$-, and $z$-direction, respectively, generating an excitation field. The calibrated vector Fluxgate (tri-axial model FLC3-70, Stefan Mayer Instruments) was positioned at the center point of the MSR. Sensor output and current monitor output of the amplifier are digitized using  a multi-channel 24-bit ADC. SF measurements were performed after degaussing with sequence \#3 (see Tab.~\ref{tab:sequences}). A typical external signal (calculated signal at the position of the sensor according to Eqn.~\ref{eqn:coilfactors} without shielding generated by one of the  Helmholtz coil pairs with AC current amplitude $I$ and frequency $f$) and the corresponding internal signal are shown in Fig.~\ref{fig:SF1}. Amplitudes and phases are extracted by fitting a sinusoidal model including an offset and linear drift, as well as up to 3rd-order harmonics to the data (harmonics generation is a result of the non-linear behavior of Mu-metal). SFs are calculated by dividing the theoretical excitation field amplitude without shielding at the sensor position (according to Eqn.~\ref{eqn:coilfactors}) by the measured absolute signal amplitude. It is interesting to note that: firstly, an an-isotropic behaviour of the MSR is possible, meaning that, an excitation in $z$-direction, for example, can lead to measurable field amplitudes not only in the $z$-sensor, but also in the $x$- and $y$-sensors. Therefore, the absolute value has been used in the denominator in the shielding value definition above. Secondly, the resulting inner field is phase shifted with respect to the the excitation field outside the MSR. Measured phase shifts are given in Fig.~\ref{fig:phases}.\\
Shielding factors do not only depend on frequency and amplitude, but also on the direction of the external signal and slightly on the magnetization status of the MSR. SFs decrease only after the MSR was massively magnetized which is a rare occurrence in practice (e.~g., using tools containing permanent magnets like electric screwdrivers inside the MSR). SFs for a wider range of parameters are given in Figs.~\ref{fig:SF2} and \ref{fig:SF3}.\\
Shielding Factors for all three directions as a function of frequency for an external amplitude of 1.75~µT are shown in Fig.~\ref{fig:SF2} (bottom data points, no shaking). SFs are constant at low frequencies up to 0.01~Hz and increase substantially with higher frequencies. At high frequencies, shielding is dominated by Eddy currents inside the highly conductive copper plated aluminum layer. Measurement uncertainties increase above 5~Hz due to the small internal signal (sensor noise limit) resulting from high SFs. Therefore, the SF for 10~Hz was measured using the single-axis Fluxgate magnetometer FL1-100 (Stefan Mayer Instruments) in the $z$-direction only.\\
Fig.~\ref{fig:SF3} shows the amplitude dependency of SFs
which increase with increasing excitation field strength. In practice, SFs at low excitation below 1~µT are most relevant as typical noise and disturbances of the ambient magnetic field are low (see Fig.~\ref{fig:residualfield}, top). A quadratic model was fitted to the data (fit results see Fig.~\ref{fig:SF3}).

%%%%%%%%%%%%%%%%%%%%%%%%%%%%%%%%
\subsection{\label{sec:shaking}Shielding Factor enhancement by low frequency, low current Mu-metal shaking}
\begin{figure}
    \centering
    \includegraphics[width=0.5\textwidth]{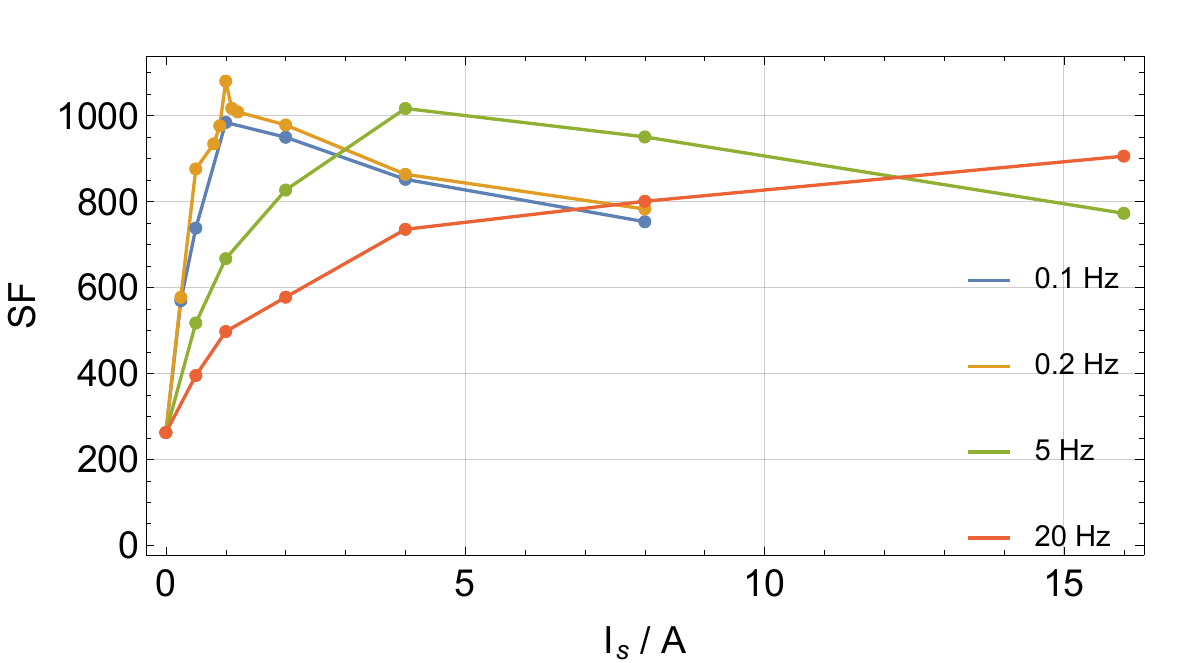}
    \caption{Shielding Factors for an external signal frequency $f=20$~mHz and amplitude $1.75$~µT in $z$-direction. Shaking of the outermost Mu-metal layer in $z$-direction was applied for a set of different shaking frequencies $f_\text{s}$ as a function of the shaking current $I_\text{s}$. Measurement uncertainties are smaller than the symbol size. Lines: guide to the eye. $I_\text{s}=0$ is equivalent to "no shaking".
    }    \label{fig:shaking}
\end{figure}

The enhancement of Shielding Factors of a high-permeability magnetic shield using "shaking", i.~e., the impression of an alternating magnetic field on the ferromagnetic material, has been known for a long time ~\cite{Spooner,Herrmannsfeldt, Cohen}. Shaking frequencies $f_\text{s}$ were typically the power line frequency (50 or 60~Hz) or even higher frequencies $f_\text{s}=400$~Hz to 1~kHz ~\cite{Sasada1, Sasada2}. The origin of this effect is not fully understood; D.~Cohen states in ~\cite{Cohen}: "One concludes that there is no obvious explanation for the shielding enhancement; apparently the [magnetic] domains need only be in motion to obtain enhancement."\\
Here, we show that SFs can be improved by a factor 4 by low frequency, low current shaking of the outermost Mu-metal layer: The signal generator, power amplifier and one of the degaussing coil units (here: $z$-direction) are used to generate the continuous alternating magnetic field inside the Mu-metal walls. SFs were measured as before using Helmholtz-like coil pairs at the outer edges of the MSR and a tri-axial Fluxgate at the center point of the MSR. Fig.~\ref{fig:shaking} shows the resulting SF with shaking applied for a set of different shaking frequencies $f_\text{s}$ as a function of the shaking current $I_\text{s}$. For a given $f_\text{s}$ one finds a SF maximum at a certain current: $I_\text{s}\approx1$~A for $f_\text{s}=0.1$ and 0.2~Hz. Optimal $I_\text{s}$ increases with increasing $f_\text{s}$ and probably lies outside the inspected range for $f_\text{s}=20$~Hz. Optimized SFs (e.~g., for $f_\text{s}=0.1$, 0.2 or 5~Hz) are a factor $\approx4$ above SFs without shaking for a wide frequency range of the external signal generated by the Helmholtz coil pairs (see. Fig.~\ref{fig:SF2}) up to 1~Hz. Above 1~Hz, they approach SFs without shaking. This is to be expected as the induction of Eddy currents in the aluminum layer is the dominant shielding mechanisms for higher frequencies, which is independent of shaking.\\
Two observations: First, one would assume that the (symmetry-breaking) choice of the shaking direction would lead to an-isotropic shielding performance. However, we found that shaking improves shielding factors in all directions almost equally. Second, shaking improves shielding performance not only below the shaking frequency $f_\text{s}$, but also above.\\
The advantages of shaking at low frequencies are: The required electrical power for shaking at $f_\text{s}=0.2$~Hz and $I_\text{s}=1$~A is only 1~W compared to e.~g. $\approx250$~W at $f_\text{s}=20$~Hz and $I_\text{s}=16$~A (see Fig.~\ref{fig:shaking}). This results in less heat deposited in the degaussing coils causing smaller temperature drifts of the surrounding Mu-metal. The smaller shaking current leads to a smaller shaking signal leakage into the interior of the MSR (which is already very small due to the closed magnetic flux loop inside Mu-metal walls). Limiting shaking to the outermost Mu-metal layer, which is the case here, effectively reduces shaking-signal leakage into the MSR, because the two inner Mu-metal layers remain for shielding the shaking signal. Measured shaking leakage at $f_\text{s}=0.2$~Hz and $I_\text{s}=1$~A at the center of the MSR is less than 8~pT (measured with a lock-in technique with an integration time of several hours). If shaking-signal leakage into the MSR is found to be of potential concern, tuning of the exact shaking frequency so as not to interfere with the actual measurement signal is possible. In our case, the Larmor frequencies of $^{129}$Xe and $^3$He are $\approx5$~Hz and $\approx12$~Hz, respectively, so that shaking leakage at $f_\text{s}=0.2$~Hz is of no concern. Furthermore, 0.2~Hz is already in the elevated $1/f$-noise region of the SQUID magnetometers~\cite{Allmendinger}.
%%%%%%%%%%%%%%%%%%%%%%%%%%%%%%%%
\subsection{\label{sec:PerformanceResidual}Residual magnetic field and drift}
\begin{figure}
    \centering
    \includegraphics[width=0.5\textwidth]{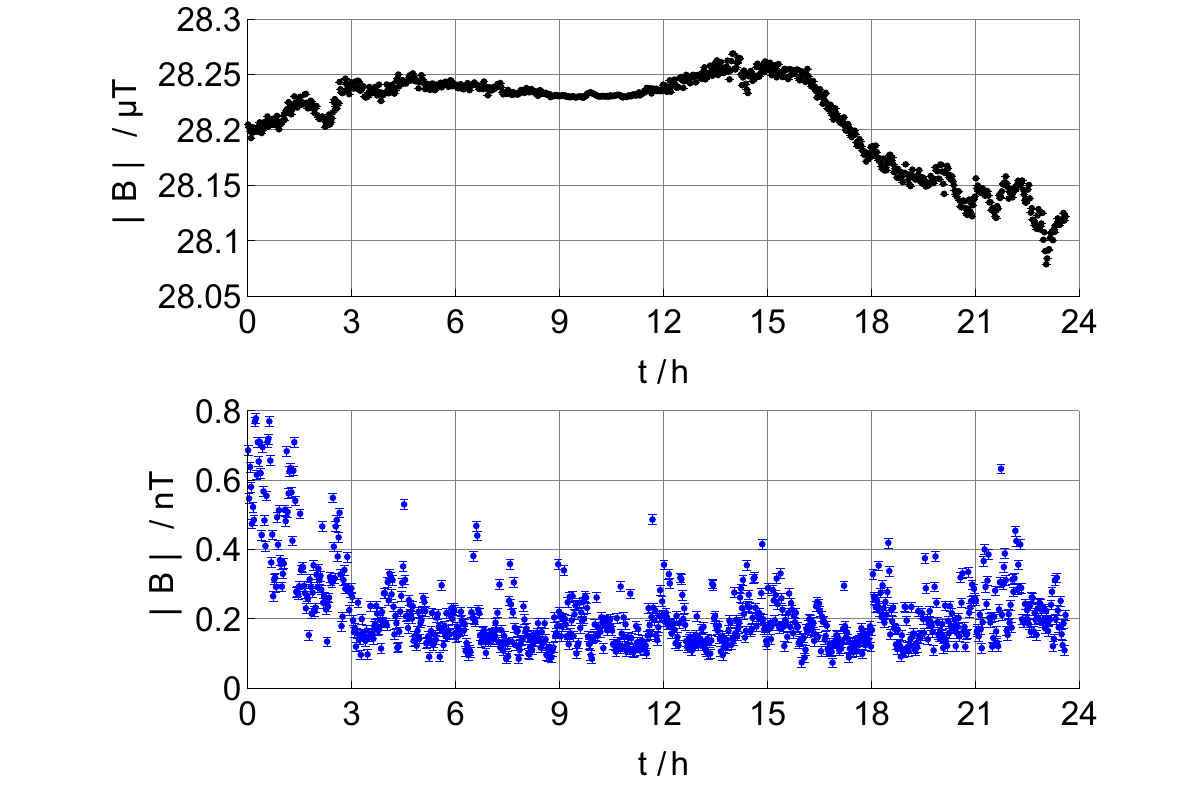}
    \caption{Magnitude of the measured external (black) and internal (blue) magnetic field, after degaussing procedure \#3 (see Tab.~\ref{tab:sequences}. Shaking at $f_\text{s}=5$~Hz and $I_\text{s}=4$~A in the $z$-direction was applied continuously. Each data point represents the average over a 100~s period. Two major observations can be stated: There is a relaxation of the residual field within the first 3~h after degaussing and no obvious correlation between the external field and the residual field inside the shield.   
    }    \label{fig:residualfield}
\end{figure}

The calibrated vector fluxgate (tri-axial model FLC3-70, Stefan Mayer Instruments) was positioned at the center point of the MSR. After leaving the door open for 30 minutes to create a reproducible magnetized state of the MSR, the door was closed and one of the different degaussing sequences according to Tab.~\ref{tab:sequences} was applied. Once the sequence was finished, the internal magnetic field was monitored for at least 24 hours. Fluctuations of the magnetic field outside the MSR were monitored by an identical fluxgate positioned 1~m in front of the center of the front wall. 
Fig.~\ref{fig:residualfield} (blue) shows the typical drift of the residual field at the center of the MSR over one day after applying degaussing procedure \#3. The residual field is slightly below 1~nT immediately after degaussing, then decreases to $\approx 200$~pT within the next three hours, and stays stable with fluctuations of less than 200~pT. This drift of the inner magnetic field within the first hours is reproducible. Most likely it is caused by a magnetic relaxation or disaccommodation~\cite{Kronmuller} of the innermost Mu-metal layer as there is no obvious correlation between the internal and external magnetic field (see Fig.~\ref{fig:residualfield}). 
The upper limits of absolute residual fields after applying the different degaussing sequences are listed in Tab.~\ref{tab:sequences}.\\
The results show, that an excellent residual magnetic field inside the MSR can be achieved with a degaussing sequence using a frequency sweep (sequence \# 3 in Tab.~\ref{tab:sequences}) in 21 minutes.

%%%%%%%%%%%%%%%%%%%%%%%%%%%%%%
\subsection{\label{sec:PerformanceGradients}Estimation of  magnetic field gradients}
\begin{figure}
    \centering
    \includegraphics[width=0.5\textwidth]{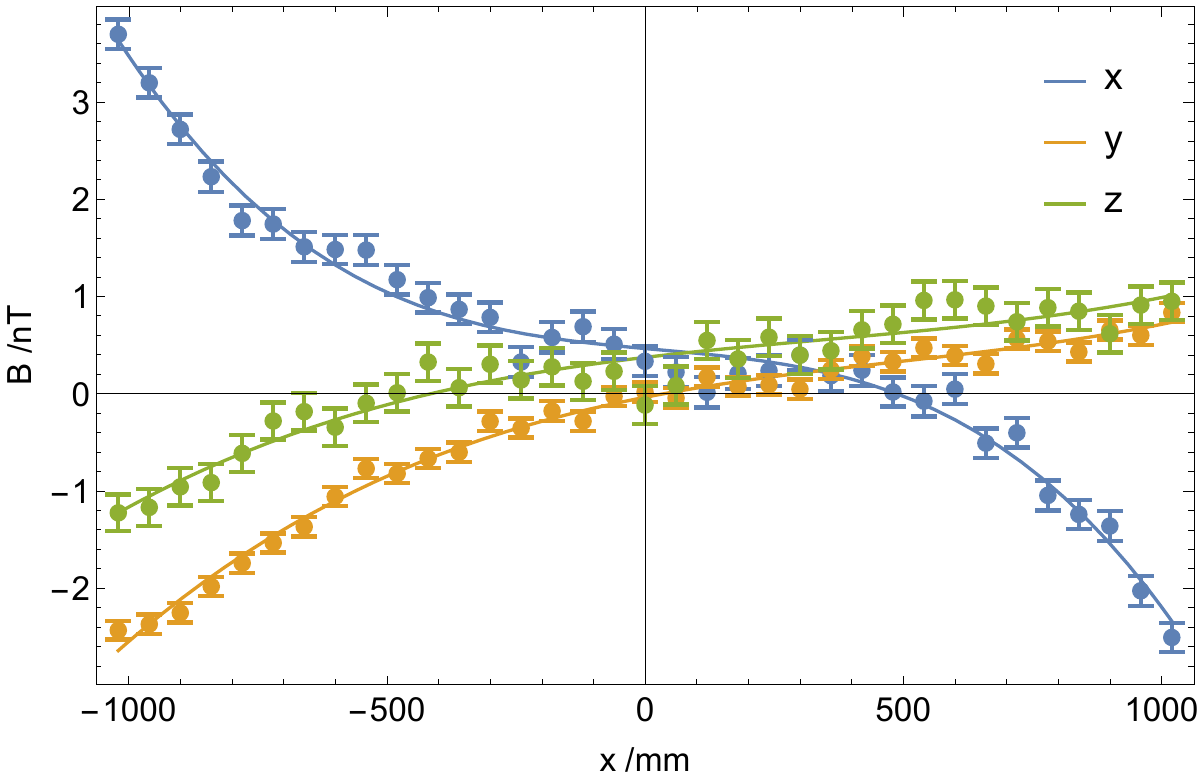}
    \caption{The three components of the residual magnetic field inside the MSR as a function of the $x$-position ($y=z=0$). Solid lines are results of 3rd-order polynomial fits. The data were taken shortly after degaussing, i.~e., within the relaxation period (see Fig.~\ref{fig:residualfield}). The total measurement time was 12~min, i.~e., short compared to the typical residual field relaxation time after degaussing.
    }    \label{fig:residualfield2}
\end{figure}

As magnetic field homogeneity strongly influences the spin coherence time and thereby the measurement sensitivity, the gradients in a central volume (200x200x200~mm$^3$) containing the spin samples are of great interest. Therefore, a vector Fluxgate (tri-axial model FLC3-70, Stefan Mayer Instruments) was mounted on a non-magnetic rail system. This allowed for residual-field measurements not only at the center of the MSR, but also over the larger span from wall to wall. We chose a straight line along the $x$-direction, i.~e., from $x= -1020$ to +1020~mm, through the center of the MSR ($y=z=0$). Magnetometer offsets were determined, and then, shortly after degaussing with sequence \# 3 (see Tab.~\ref{tab:sequences}), the sensor was moved in steps of 60~mm by pulling on strings which were guided outwards via pulleys. Immediately after the measurement process was completed, magnetometer offsets were determined again. The field components as a function of $x$ are shown in Fig.~\ref{fig:residualfield2}. Magnetic field gradients were determined by taking the derivatives with respect to $x$ of 3rd-order polynomial fits to the measurement data. The magnitude of the individual gradients was less than 10~pT/cm in the central volume. Note that only three of the five independent magnetic gradients were determined; however, all are expected to be on the same order of magnitude due to the symmetry of the MSR. The field homogeneity achieved is high enough so that transverse spin relaxation times $T_2^*$ of many hours will be reached under measurement conditions described, e.~g., in ~\cite{Gemmel}.\\

 %%%%%%%%%%%%%%%%%%%%%%%%%%%%%%%%%%%%%%%%%%%%%%%%%%%%%%%%%%%%%%%%%%%%%%%%%%%%%
 %%%%%%%%%%%%%%%%%%%%%%%%%%%%%%%%%%%%%%%%%%%%%%%%%%%%%%%%%%%%%%%%%%%%%%%%%%%%%
\section{\label{sec:Motivation}Motivation for the chosen degaussing procedure}
In this section, we motivate the unusual choice of our degaussing procedure (especially the low frequency of 0.5~Hz and the frequency sweep, see Fig.~\ref{fig:waveform}) by means of an online hysteresis measurement of the assembled MSR and Eddy-current simulations. The measured material properties are used as input parameters for the simulations.
\subsection{\label{sec:hysteresis}Hysteresis measurements}
\begin{figure}
    \centering
    \includegraphics[width=0.5\textwidth]{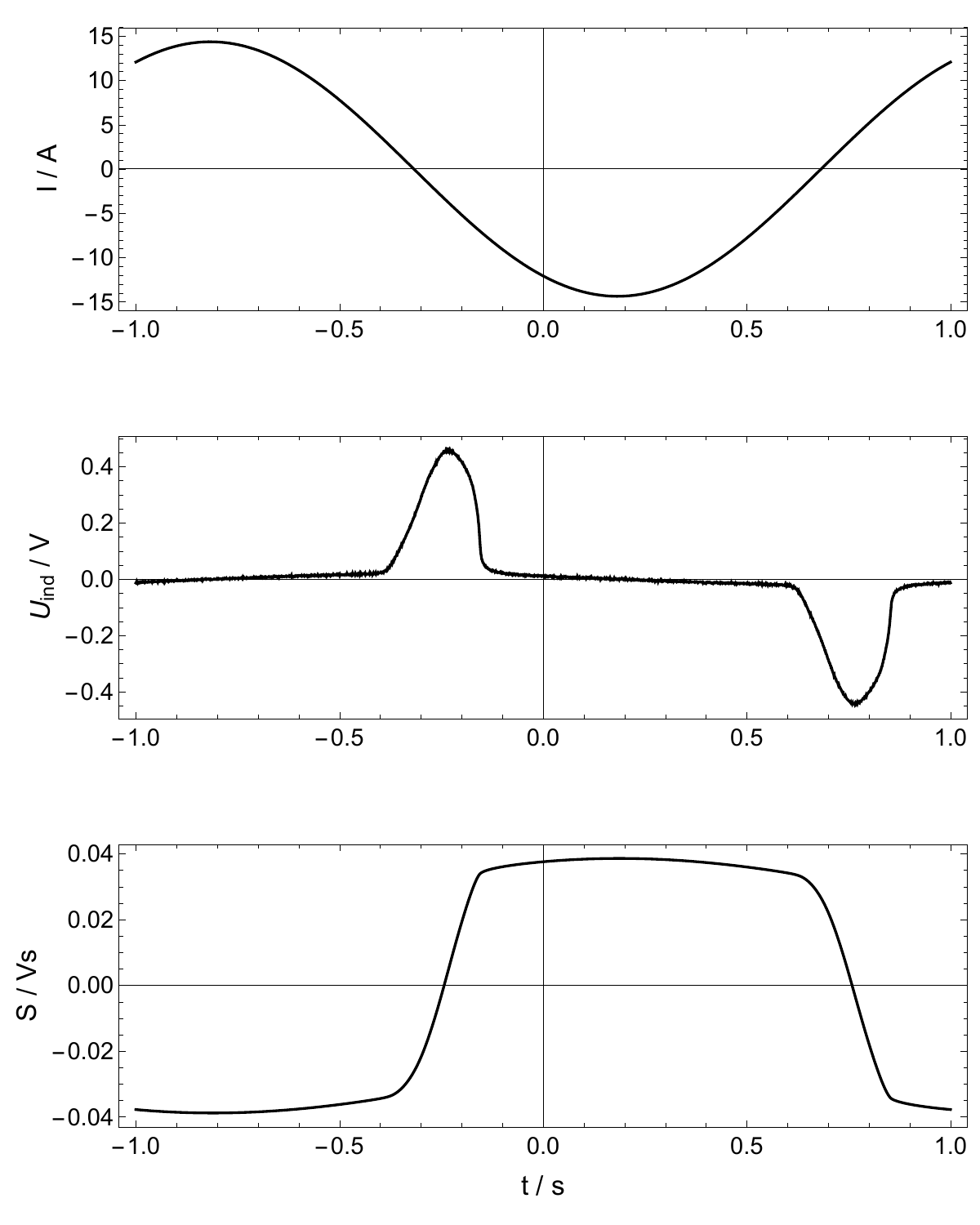}
    \caption{Hysteresis measurement using excitation coil with current $I(t)$ of sinusoidal form (here $f=0.5$~Hz (top)); measured induction voltage $U_\text{ind}$ at the pickup coil (middle); and induction voltage integral (or sum) $S$ (bottom). Averaging over several periods is applied to reduce noise. 
    }    \label{fig:hyst_principle}
\end{figure}

\begin{figure}
    \centering   
    \includegraphics[width=0.5\textwidth]{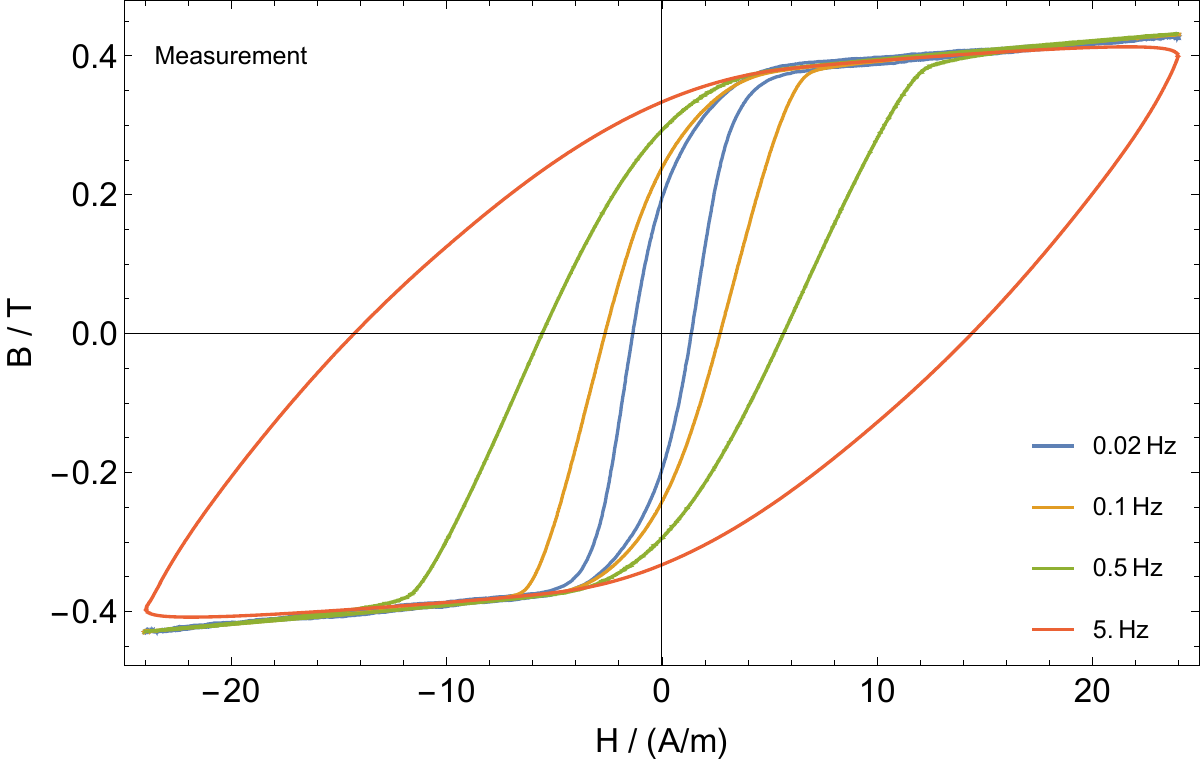}
    \includegraphics[width=0.5\textwidth]{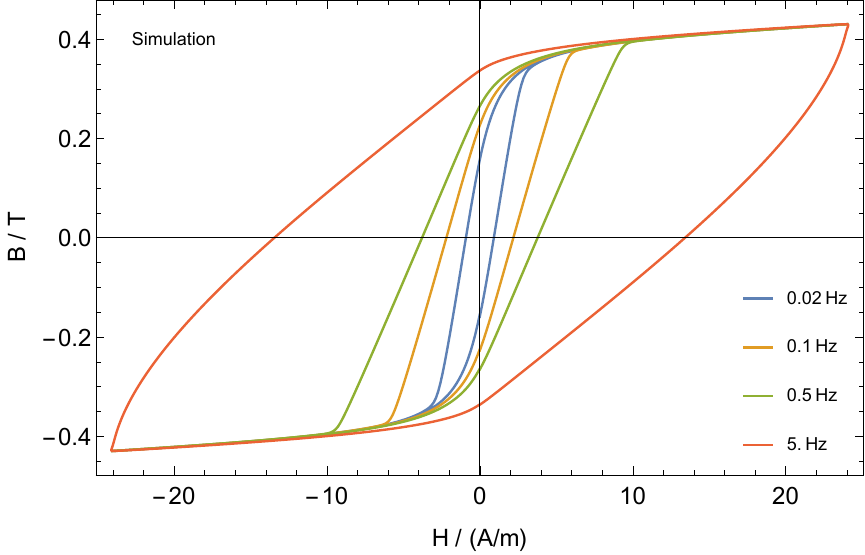}
    \caption{Top: Hysteresis measurement results at the assembled MSR walls (outer layer) for a set of different frequencies.
    Bottom: Simulated hysteresis curves (see Sec.~\ref{sec:simulation})
    }    \label{fig:hyst_result}
\end{figure}

The material properties, especially the magnetic permeability $\mu_r$, of Mu-metal in the assembled MSR are of great interest, as mechanical stress during assembly might lead to a degradation of the magnetic properties, e.~g., reduced $\mu_r$. Measuring the magnetic hysteresis curve, i.~e., the magnetic flux density $B$ vs. the magnetic field $H$ gives access not only to $\mu_r$, but also to a possible remanence. The degaussing setup (see Fig.~\ref{fig:degcoils}) can be used to measure the magnetic hysteresis curve of the Mu-metal of the fully assembled MSR assuming homogeneous material properties (wall thickness and permeability, for example) by the following technique:\\
The magnetic field inside the closed flux loop (as depicted in Fig.~\ref{fig:degcoils}) with total length $l=4\cdot3$~m embracing the four walls is given by
\begin{eqnarray}
H=\frac{N_1\cdot I}{l}~~.
\label{eqn:hyst_Hfield}
\end{eqnarray}
 Here, $N_1=20$ is the  winding number of the excitation coil (=degaussing coil unit), carrying a current of $I$.
 A pickup coil with winding number $N_2=10$ was placed along the edge connecting two walls of the outer Mu-metal layer. A high impedance measurement of the induction voltage $U_\text{ind}$ gives access to the magnetic flux density $B$ via 
\begin{eqnarray}
U_\text{ind}&=&-N_2\frac{d\Phi}{dt}=-N_2A\frac{dB}{dt}\nonumber \\
\Rightarrow B&=&-\frac{1}{N_2A}\int U_\text{ind}\,dt+C~.
\label{eqn:hyst_Uind}
\end{eqnarray}
Here, $A=3\cdot0.003~\text{m}^2$ is the cross section of the Mu-metal wall covered by the pickup coil. The integration constant $C$ is of no further importance and will be adjusted later so that the hysteresis curve is symmetrical around $B=0$. In practice, $U_\text{ind}$ is sampled with a frequency $f_\text{ADC}$, so that the integral in Eqn.~\ref{eqn:hyst_Uind} can be approximated using
\begin{eqnarray}
S=\int U_\text{ind}\,dt \approx&\sum_{i} U_\text{ind}(t=i/f_\text{ADC})\cdot \frac{1}{f_\text{ADC}}~.
\label{eqn:hyst_S}
\end{eqnarray}
The measurement process is as follows (see Fig.~\ref{fig:hyst_principle}): A sinusoidal signal with frequency $f$ is fed to the power amplifier that delivers a proportional current with a typical amplitude of $I_0=15$~A to the excitation coil. The current-monitor output of the power amplifier and the pickup-coil voltage are monitored using a 18-bit ADC with adjustable sampling frequency (typically $f_\text{ADC}=10$~kHz). The current and induction-voltage data are averaged over several periods (typically 10 to 100) to reduce noise. Then the voltage integral $S$ is calculated using Eqn.~(\ref{eqn:hyst_S}). Typical $I$, $U_\text{ind}$ and $S$ curves for $f=0.5$~Hz are shown in Fig.~\ref{fig:hyst_principle}.  Using Eqns.~(\ref{eqn:hyst_Hfield}) and (\ref{eqn:hyst_Uind}), and plotting $B(t)$ vs. $H(t)$ gives the hysteresis curve (see Fig.~\ref{fig:hyst_result}, top). For low frequencies $f=0.02$~Hz to $f=0.5$~Hz, one can easily identify the two branches for increasing and decreasing $H$ and the saturation value at $B\approx 0.4$~T. The slope gives the magnetic permeability $\mu_0\mu_r$ with $\mu_r\approx 2.5\cdot10^5$ at the steepest part of the narrow $f=0.02$~Hz curve.
The strongly non-linear relation between the magnetic flux density $B$ and the magnetic field $H$ inside the Mu-metal can be modelled according to:
\begin{align}
	B(H) = \alpha \frac{2}{\pi} \arctan( \beta H) +\gamma H
        \label{eqn:BH2}
\end{align}
with the three free parameters $\alpha$ (saturation), $\beta$ (linked to $\mu_r$) and $\gamma$ (a linear contribution). Note that, here, we use a purely phenomenological model with no remanence (Mu-metal is very soft-magnetic) with the sole purpose of extracting the relevant material properties that will be used as input parameters for Eddy-current simulations covered in the next section. From hysteresis measurements of the MSR walls at very low frequencies ($f=0.01$~Hz) and $H$ between $\approx-24$ and $+24$~A/m, we extracted the parameters $\alpha=0.4$~T, $\beta=1.26$~m/A and $\gamma=1.6\cdot10^{-3}$~Tm/A. Note that the model according to Eqn.~(\ref{eqn:BH2}), especially the linear contribution, is only valid for small magnetic fields $|H|<24$~A/m. The linear increase was also observed in ~\cite{Sun2} for a similar magnetic field of $H\approx20$~A/m.\\
For increasing frequency, the curves become broader and approach ellipses. The reason lies in the induction of Eddy currents within the 3~mm thick Mu-metal walls which effectively shield the internal material, i.~e., the resulting magnetic field at the center of the Mu-metal walls is reduced, with the consequence that saturation is not reached. Details and a numerical simulation of this effect will be covered in the next section.\\
In conclusion, our investigations have shown that the shape of measured hysteresis curves of a given high-permeability material depends strongly on the measurement frequency (see Fig.~\ref{fig:hyst_result}) and on the sample geometry (especially the material thickness). Hysteresis-curve measurements of assembled MSRs with a wall thickness $>1$~mm should be performed with low frequencies $f<0.1$~Hz. In our case we have extracted a $\mu_r\approx 2.5\cdot10^5$ of the assembled Mu-metal layer.
%%%%%%%%%%%%%%%%%%%%%%%%%%%%%%%%%%%%%%%%%
\subsection{\label{sec:simulation}Eddy-current simulation}
\begin{figure}
    \centering
    \includegraphics[width=0.5\textwidth]{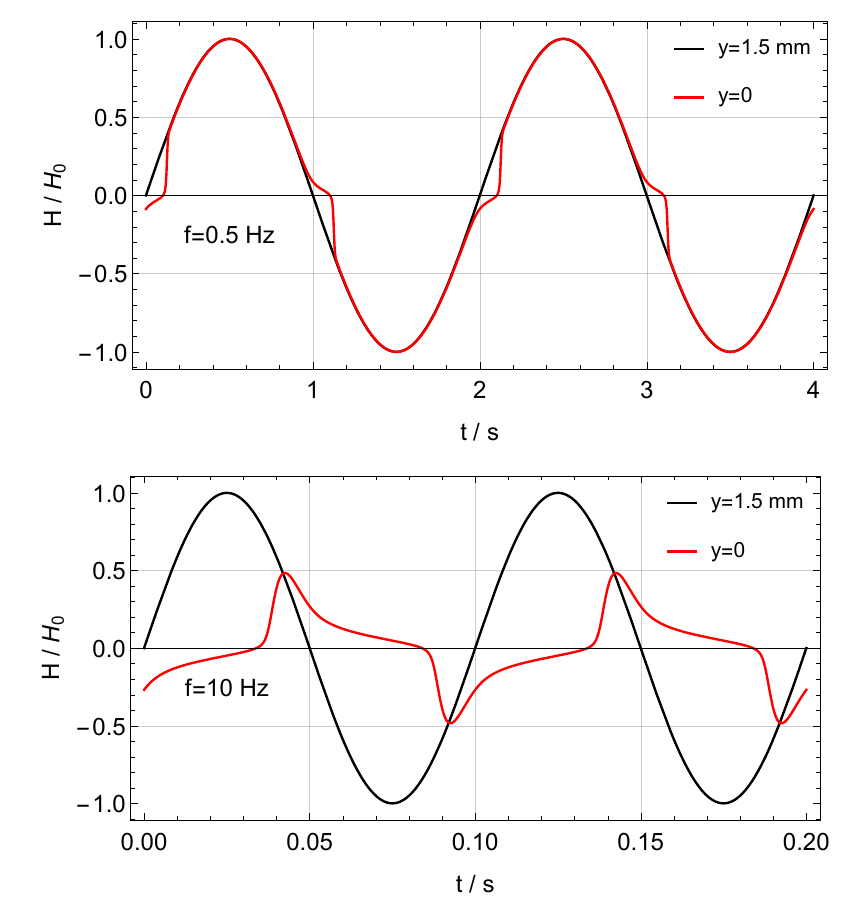}
    \caption{Eddy-current simulation: Magnetic field $H$ (normalized to the amplitude of the external field $H_0=24.1$~A/m) at the surface of the Mu-metal layer ($y=1.5$~mm, black curve) and at the core of the Mu-metal layer ($y=0$, red curve)  for $f=0.5$~Hz (top) and $f=10$~Hz (bottom).
    }    \label{fig:Ht_inside_mumetall}
\end{figure}

\begin{figure}
    \centering
    \includegraphics[width=0.5\textwidth]{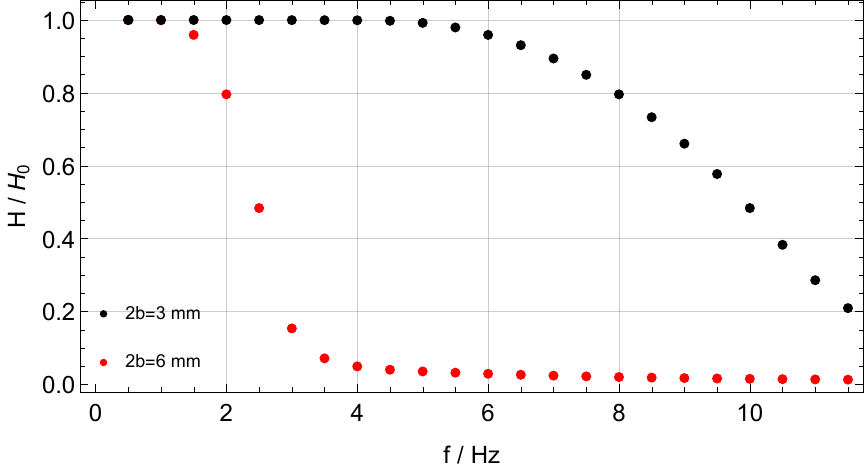}
    \caption{Eddy-current simulation: Maximum of the magnetic field $H$ (normalized to the amplitude of the external field $H_0=24.1$~A/m) at the core of the Mu-metal layer ($y=0$)  as a function of frequency $f$  for the actual wall thickness $2\cdot b=3$~mm (black), and for $2\cdot b=6$~mm (red).
    }    \label{fig:H_vs_f}
\end{figure}

\begin{figure}
    \centering
    \includegraphics[width=0.5\textwidth]{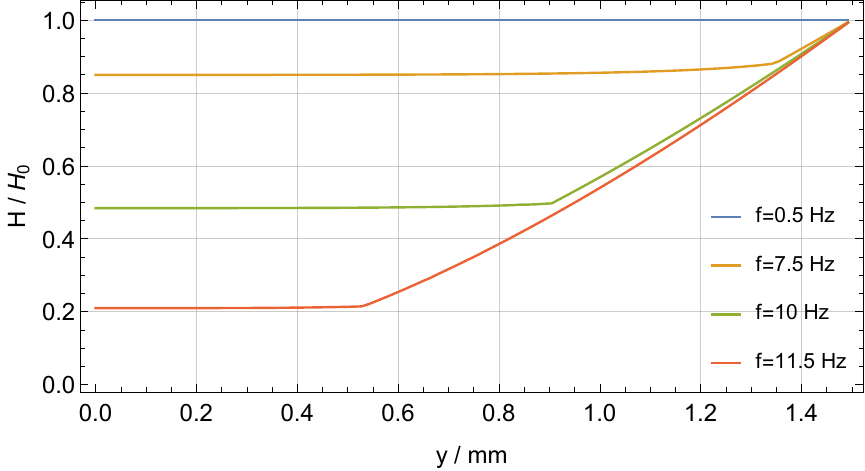}
    \caption{Eddy-current simulation: Maximum of the magnetic field $H$ (normalized to the amplitude of the external field $H_0=24.1$~A/m) as a function of the position $y$ inside the Mu-metal layer ($2\cdot b=3$~mm) for a set of different frequencies $f$.
    }    \label{fig:H_vs_y}
\end{figure}

We wish to investigate the effective shielding of internal material within the Mu-metal wall (actual thickness: 3~mm) caused by Eddy currents.\\
The thickness of the Mu-Metal wall is small compared to its length and width. For our simulation, we consider a Mu-metal plate parallel to the $xz$-plane with infinite height and infinite length and finite thickness $2b=3$~~mm in $y$-direction (spanning from $y=-b$ to $y=+b$). Note that, here, we defined a coordinate system which differs from the one introduced at the beginning (Fig.~\ref{fig:degcoils}). The degaussing coils in our simulation now produce an external homogeneous oscillating  field in $z$-direction of
\begin{align}
	\vec{H}(\vec{x}, t) = H_0 \sin(2 \pi f t) \vec{e_z}~.
 \label{eqn:Ht}
\end{align}
As the frequency $f$ of the oscillating field is very low, we can use the quasi-static Maxwell equations and Ohm's law:
\begin{align}
	\nabla \times \vec{H} &= \vec{j} \label{eqn:maxwell1}\\ 
	\nabla \times \vec{E} &= -\partial_t \vec{B} \label{eqn:maxwell2}\\ 
	\nabla \cdot \vec{B} &= 0 \label{eqn:maxwell3}\\ 
	\vec{j} &= \sigma \vec{E}~~. \label{eqn:maxwell4}
\end{align}
with current density $\vec{j}$, electric field $\vec{E}$, and conductivity $\sigma$.
Furthermore, we need the relation between the magnetic flux density $B$ and the magnetic field $H$ inside the Mu-metal according to Eqn.~(\ref{eqn:BH2}).\\
Taking the curl ($\nabla \times$) of Eqn.~(\ref{eqn:maxwell1}) and using the \textit{curl-of-the-curl} identity, then applying  Eqns.~(\ref{eqn:maxwell4}) and (\ref{eqn:maxwell2}) we obtain:
\begin{align}
\nabla \left(\nabla \cdot \vec{H}\right) - \Delta \vec{H} &= -\sigma \partial_t (\vec{B}(\vec{H}))
 \label{eqn:PDE1}
 \end{align}
We aim to find a solution to the set of differential Eqns.~(\ref{eqn:PDE1}) and (\ref{eqn:BH2}) with the boundary condition in Eqn.~(\ref{eqn:Ht}).
To do so, we take advantage of the symmetry of the problem: As the Eddy currents always point in the $x$-direction and $\vec{H}$ always points in the $z$-direction, one can reduce the problem to one dimension:
\begin{align}
	\vec{H}=\left(0,0,h(y,t)\right)~~.
\end{align}
Then, the partial differential Eqn.~(\ref{eqn:PDE1}) can be simplified to
\begin{align}
	\partial_y^2 h(y,t) = \sigma \partial_t(h(y,t)) B^{\prime}(h(y,t))
 \label{eqn:PDE2}
\end{align}
with the boundary conditions
\begin{align}
	h(-b, t) &= h(b, t) = H_0\sin(2\pi f t) \\
	h(y, 0) &= 0~~.
\end{align}
Here, $B^{\prime}$ is the derivative of $B$ with respect to its argument. The input parameters are the material properties of Mu-metal: the electrical conductivity $\sigma=8.72\cdot10^5$~S/m ~\cite{Vac1}, and the $B(H)$-dependency, modelled according to Eqn.~(\ref{eqn:BH2}) with the free parameters $\alpha=0.4$~T, $\beta=1.26$~m/A and $\gamma=1.6\cdot10^{-3}$~Tm/A, extracted from hysteresis measurements (see Sec.~\ref{sec:hysteresis}).\\
Numerical solutions of the differential equation were found using the FEM-Solver of Wolfram Mathematica on a rectangular 200x5000 grid in $y$- and $t$-direction. Simulated "hysteresis loops" can be extracted by numerically integrating the flux density in space to get the magnetic flux "seen" by the pick up coil (see Fig.~\ref{fig:hyst_result}, bottom). Simulation results reproduce the measured hysteresis loops for a broad range of frequencies. We therefore conclude, that our Mu-metal model is accurate enough for the intended purpose of studying the effect of the Eddy-current shielding of the Mu-metal core. Note that the hysteresis loop broadening with higher frequencies is solely a result of induced Eddy currents as the  $B(H)$ dependency in Eqn.~(\ref{eqn:BH2}) models Mu-metal as completely soft-magnetic.\\
Fig.~\ref{fig:Ht_inside_mumetall} shows the magnetic field $H$ normalized to the excitation field amplitude $H_0=24.1$~A/m as a function of time at the center ($y=0$, red curve) and at the surface ($y=1.5$~mm, black curve) of the 3~mm thick Mu-metal wall for $f=0.5$~Hz (top) and 10~Hz (bottom). One observes that for $f=10$~Hz, the maximum of the internal field ($y=0$) is reduced to $H/H_0\approx0.5$ (and shifted in time), while for the lower frequency no relevant reduction of the internal field is observed. Fig.~\ref{fig:H_vs_f} shows this effect, i.~e., the maximum normalized field at the Mu-metal core, for a broader range of frequencies for the actual wall thickness $2\cdot b=3$~mm (black), and for $2\cdot b=6$~mm (red). Doubling the Mu-metal wall thickness reduces the frequency where $H/H_0=0.5$ from 10 to 2.5~Hz. Fig.~\ref{fig:H_vs_y} shows that (for $2\cdot b=3$~mm) the maximum $H$ is not only reduced at the center but also in a larger central volume, e.~g., for $f=10$~Hz $H/H_0\approx0.5$ for more than half of the Mu-metal volume ($y$ from -0.9 to 0.9~mm).\\
The Eddy-current simulations have the following possible limitations: The assumption of infinite Mu-metal walls in two dimensions is a good approximation (3~mm wall thickness compared to the 3~m wall height and length, furthermore four walls generate a closed loop flux path); however, the edges might behave differently. Secondly, we assumed homogeneous material properties; variations in material  thickness (due to small overlaps), permeability, or conductivity (which might be smaller between laminated sheets) are not accounted for. Finally, for a deeper understanding of the degaussing process, the magnetic properties of Mu-metal have to be described by a more detailed model: e.~g., the  frequency-dependent \textit{Jiles-Atherton} model ~\cite{Jiles1, Jiles2,Jiles3} considering the energy changes of magnetic domains during magnetization (or degaussing), or the computation-time efficient \textit{phase-shift} approach ~\cite{Sun2}.
\subsection{\label{sec:consequences}Consequences for degaussing and shaking method}
For effective degaussing leading to a reproducible low residual field inside the MSR, the maximum peak applied magnetic field must be sufficiently high to saturate the magnetic material in every region of the shielding. This calls for a low degaussing frequency to saturate the innermost material. We chose $f=0.5$~Hz to have a large safety margin accounting for edge effects and material in-homogeneity like variations in thickness (due to small overlaps) or permeability. At $f=0.5$~Hz saturation is reached at the core of the Mu-metal even for a wall thickness of 6~mm (see Fig.~\ref{fig:H_vs_f}). Note that for degaussing effectively at $f=0.5$~Hz, $H_0\approx24$~A/h is a sufficiently large initial magnetic field strength. The reason is that for $f=0.5$~Hz the two branches of the hysteresis curve coincide in the region $|H|>15$~A/m (see Fig.~\ref{fig:hyst_result}).\\
Then, the amplitude decrease (see last section in Fig.~\ref{fig:waveform}) must be slow enough, so that consecutive maxima have a small difference (ratio of consecutive maxima $r\approx0.99$), leading to a random orientation of magnetic domains. This is usually achieved by slowly decreasing the current through the degaussing coils which leads to long time constants at low degaussing frequencies and therefore long total duration (see \#1 in Tab.~\ref{tab:sequences}).\\
An alternative can be deduced from Fig.~\ref{fig:H_vs_f}:
Increasing the frequency of the degaussing current while the amplitude stays constant leads to decreasing resulting fields in the center of the Mu-metal. We chose a linear frequency increase (sweep) from $f=0.5$~to 10~Hz within $t_\text{sweep}=100$~s. Subsequently, $f=10$~Hz stays constant and the current amplitude decreases exponentially. Due to the higher frequency, the time constant is substantially shorter for a given $r$. Together with time-saving modifications of the degaussing sequence and $r$, a reproducible residual magnetic field below 1~nT can be achieved in 21 minutes (see \#3 in Tab.~\ref{tab:sequences}).\\
Similarly, the Eddy-current simulations motivate shaking at low frequencies. A low-frequency shaking field penetrates to the core of the Mu-metal layer leading to increased shielding performance, while the shaking current can be drastically reduced (see Fig.~\ref{fig:shaking}).
%%%%%%%%%%%%%%%%%%%%%%%%%%%%%%%%%%%%%%%%%%%
%%%%%%%%%%%%%%%%%%%%%%%%%%%%%%%%%%%%%%%%%%%
\section{\label{sec:Summary}Summary and Outlook}
We described the properties of the Magnetically Shielded Room (MSR), manufactured by Vacuumschmelze, intended for next level $^3$He/$^{129}$Xe co-magnetometer experiments which require improved magnetic conditions. The degaussing (magnetic equilibration) procedure was improved for the 3~mm thick Mu-metal layers. The key is reaching saturation at the center of the Mu-metal layer by using a low initial degaussing frequency while reducing the necessary time for degaussing using a frequency sweep with constant amplitude followed by an exponential decay of the amplitude. Degaussing parameters were found by online hysteresis measurements and by Eddy-current simulations. The investigations have shown that Eddy currents have to be taken into account in hysteresis-curve measurements of high-permeability material like Mu-metal. Such measurements, e.~g., on assembled MSRs with a wall thickness $>1$~mm, have to be performed with low frequencies $f<0.1$~Hz. Using higher frequencies causes a substantial broadening of the measured hysteresis curve and might lead to a false interpretation of the magnetic properties. In our case we have extracted a remarkably high permeability $\mu_r\approx 2.5\cdot10^5$ of the assembled Mu-metal layer. The degaussing procedure for the whole MSR takes 21 minutes and measurements of the residual magnetic field using Fluxgate magnetometers show that $|B|<$~1nT can be reached reliably. Shielding Factors can be improved by a factor $\approx 4$ in all directions by low frequency (0.2~Hz), low current (1~A) shaking of four walls of the outermost Mu-Metal layer.\\
Determining the properties of the residual field inside the MSR, especially gradients and their temporal stability, is essential for the intended use of the MSR for $^3$He/$^{129}$Xe co-magnetometer experiments. As a first estimate, we found that the order of magnitude of the gradients is 10~pT/cm in the central volume. However, noise and offset drift of Fluxgate magnetometers quickly limit the precision that can be achieved with the method above. Consequently, we intend to use the following method: Magnetic field gradients can be extracted very precisely and accurately from transverse relaxation rates of precessing spin samples, e.~g., gaseous, nuclear spin polarized $^3$He and $^{129}$Xe atoms in a spherical cell, which was demonstrated in previous work~\cite{Allmendinger3}, where a resolution below pT/cm was reached. Finally, gradient compensation by a coil system is intended: systematically adjusting the gradient coil currents and simultaneously monitoring and minimizing transverse relaxation rates according to a downhill simplex algorithm~\cite{Allmendinger} leads to minimized magnetic field gradients in the volume of interest which in this case is the volume occupied by the spin sample.\\
%%%%%%%%%%%%%%%%%%%%%%%%%%%%%%%%%%%%%%%%%%%%%%%%%%%%%%%%%%%%%%%%%%%%%%%%%%%%%%%%%%%%%%%%%%%%%%%%%%%%%%%%
\begin{acknowledgments}
We acknowledge the excellent support during the planing and construction phase of the MSR provided by L.~Bauer, M.~Hein, M.~W\"ust, M.~Staab and J.~Gerster of VACUUMSCHMELZE GmbH, Germany, and are grateful for fruitful discussions about amplitude-dependent shielding factors. We thank H.-J.~Krause, FZ J\"ulich and S. Hummel, PI Heidelberg for their technical support. This work was supported by the Deutsche Forschungsgemeinschaft (DFG, German Research Foundation), Grant No. 455449148.\\
\end{acknowledgments}
\section*{Author Declarations}
\subsection*{Conflict of Interest}
The authors have no conflicts to disclose.\\
\subsection*{Author contributions}
{\bf Fabian Allmendinger}: Conceptualization (lead); methodology (lead); investigation (lead); formal analysis (lead); writing – original draft (lead); writing – review and editing (equal).
{\bf Benjamin Brauneis}: Methodology; investigation; software; writing – review and editing (equal).
{\bf Werner Heil}: Methodology; writing – review and editing (equal).
{\bf Ulrich Schmidt}: Supervision; methodology; writing – review and editing (equal).

\section*{Data Availability Statement}
The data that support the findings of this study are available from the corresponding author upon reasonable request.

\nocite{*}
\bibliographystyle{unsrt}
\bibliography{mainbib}% Produces the bibliography via BibTeX.

\end{document}